\documentclass{article}
\usepackage{graphicx}
\usepackage{amsmath}
\usepackage{amssymb}
\usepackage{threeparttable}
\usepackage{braket}
\usepackage{siunitx}

\newcommand\given[1][]{\:#1\vert\:}
\newcommand{\Mdata}{M_{\text{data}}}
\newcommand{\KS}{K_{\text{S}}}
\newcommand{\bbQ}{{\boldsymbol{Q}}}
\newcommand{\bbqRes}{ \text{\bf q} }
\newcommand{\bvec}[1]{\boldsymbol{#1}}
\newcommand{\myeqref}[1]{Eqn.~\eqref{#1}}

\DeclareMathOperator*{\argmin}{arg\,min}
\title{An encryption-decryption framework to validating single-particle imaging}

\usepackage{authblk}
\author[a,b]{Zhou Shen}
\author[a,b]{Colin Zhi Wei Teo}
\author[c,d]{Kartik Ayyer}
\author[a,b,e]{N. Duane Loh}
\affil[a]{Centre for Bio-imaging Sciences, National University of Singapore, 14 Science Drive 4, 117557, Singapore}
\affil[b]{Department of Physics, National University of Singapore, 2 Science Drive 3, 117551, Singapore}
\affil[c]{Max Planck Institute for the Structure and Dynamics of Matter, Luruper Chaussee 149, 22761 Hamburg, Germany}
\affil[d]{Center for Free-Electron Laser Science, Luruper Chaussee 149, 22761 Hamburg, Germany}
\affil[e]{Department of Biological Sciences, National University of Singapore, 14 Science Drive 4, 117557, Singapore}
\date{}

\begin{document}
\maketitle
\begin{abstract} 
We propose an encryption-decryption framework for validating diffraction intensity volumes reconstructed using single-particle imaging (SPI) with x-ray free-electron lasers (XFELs) when the ground truth volume is absent. 
This framework exploits each reconstructed volumes' ability to decipher latent variables (e.g.~orientations) of unseen {\it sentinel} diffraction patterns.
Using this framework, we quantify novel measures of orientation {\it disconcurrence}, inconsistency, and disagreement between the decryptions by two independently reconstructed volumes.
We also study how these measures can be used to define data sufficiency and its relation to spatial resolution, and the practical consequences of focusing XFEL pulses to smaller foci. 
This framework overcomes critical ambiguities in using Fourier Shell Correlation (FSC)\cite{Harauz1986-nd} as a validation measure for SPI.
Finally, we show how this encryption-decryption framework naturally leads to an information-theoretic reformulation of the resolving power of XFEL-SPI, which we hope will lead to principled frameworks for experiment and instrument design.
\end{abstract}
\section{Introduction}
X-ray free-electron lasers (XFELs) are a promising tool for studying the three-dimensional (3D) structures of macromolecular assemblies \cite{Spence2017-dz, Chapman2019-wz}. 
The short and intense XFEL pulses make it possible to collect diffraction patterns of a macromolecule before the XFEL-damaged atomic nuclear motions become substantial \cite{Neutze2000-mf, Jurek2004-lm, Chapman2006-fl, Yoon2016-mu, Fortmann-Grote2017-iq}.

XFEL pulses are sufficiently intense and coherent for single-particle imaging (SPI), where a single macromolecule can scatter enough photons for us to infer its 3D orientation, hence structure \cite{Loh2009-eg, ayyer2016dragonfly, Kassemeyer2013-yx, Yoon2011-vc}.
XFEL-SPI makes the difficult task of growing large, well-diffracting macromolecular crystals (even micrometer size ones \cite{Chapman2011-la}) unnecessary.

\begin{figure}[ht]
    \centering
    \includegraphics[width=0.7\textwidth]{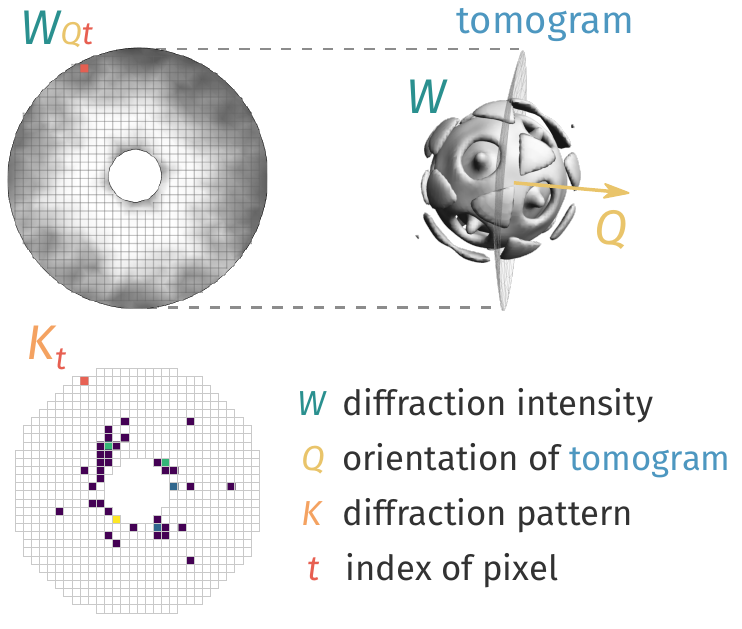}
    \caption{Schematic of how orientations are encoded in XFEL-SPI. A diffraction pattern collected on a detector ($K_t$ where $t$ labels the pixels on the detecor) of a scatterer is an Ewald tomogram $W_{Qt}$ through the 3D diffraction volume $W$. When this scatterer suffers an active random 3D rotation $\Omega$ about its own original reference frame, it is equivalent to a passive rotation of said Ewald tomogram in the opposite sense (i.e.~$\Omega^{-1}$). Throughout the rest of the paper, we parametrize this rotation with unit quaternions $Q \equiv \Omega(Q)$ (primer on unit quaternions in Appendix).}
    \label{fig:rotation_schematic}
\end{figure}

Instead, desiccated samples are randomly injected into a regular train of XFEL pulses with random orientations. 
To understand how orientations are defined in SPI, consider what happens when a scatterer, whose 3D diffraction volume is denoted $W$, is presented to the SPI laboratory reference frame (Fig.~\ref{fig:rotation_schematic}).

Collected diffraction patterns are identified and analyzed in various ways including: 
determining the 3D structures that most likely produced the ensemble of SPI patterns \cite{ekebergThreeDimensionalReconstructionGiant2015}, or studying the range of 3D morphologies spanned by the XFEL scatterers \cite{Loh2012-zx, Van_der_Schot2015-fk, Hantke2014-kb}.

Reconstructing a set of 3D structure from many SPI patterns comprises three sequential stages, each of which can be considered for validation \cite{Yoon2016-mu}.
These stages are: recovering a 3D diffraction intensities $W$ from many two-dimensional (2D) SPI patterns; 
using phase-retrieval to reconstruct the 3D realspace scattering density from $W$; 
fitting atomic coordinates to the scattering density. 
Separate validation routines between these stages can help diagnose where resolution loss might have occurred.

This work focuses on validating the first stage, where we reconstruct $W$ by inferring the latent 3D orientations of SPI diffraction patterns.
This inference is challenging for small macromolecules that produce weak diffraction patterns. 
In these cases, the Fourier Shell Correlation (FSC) \cite{Harauz1986-nd}, which is typically used to validate 3D structures recovered using cryo-electron microscopy, has become increasingly popular for estimating spatial resolution\cite{hantkeHighthroughputImagingHeterogeneous2014,xuSingleshotThreedimensionalStructure2014,ekebergThreeDimensionalReconstructionGiant2015,ayyerLowsignalLimitXray2019,giewekemeyerExperimental3DCoherent2019,hosseinizadehConformationalLandscapeVirus2017,
ikonnikovaReconstruction3DStructure2019,kimReconstruction3DImage2020,nakanoSingleparticleXFEL3D2018,
poudyalSingleparticleImagingXray2020,pryorSingleshot3DCoherent2018,roseSingleparticleImagingSymmetry2018,
shiEvaluationPerformanceClassification2019,vonardenneStructureDeterminationSingle2018}. 

\begin{figure}[h!]
    \centering
    \includegraphics[width=0.6\textwidth]{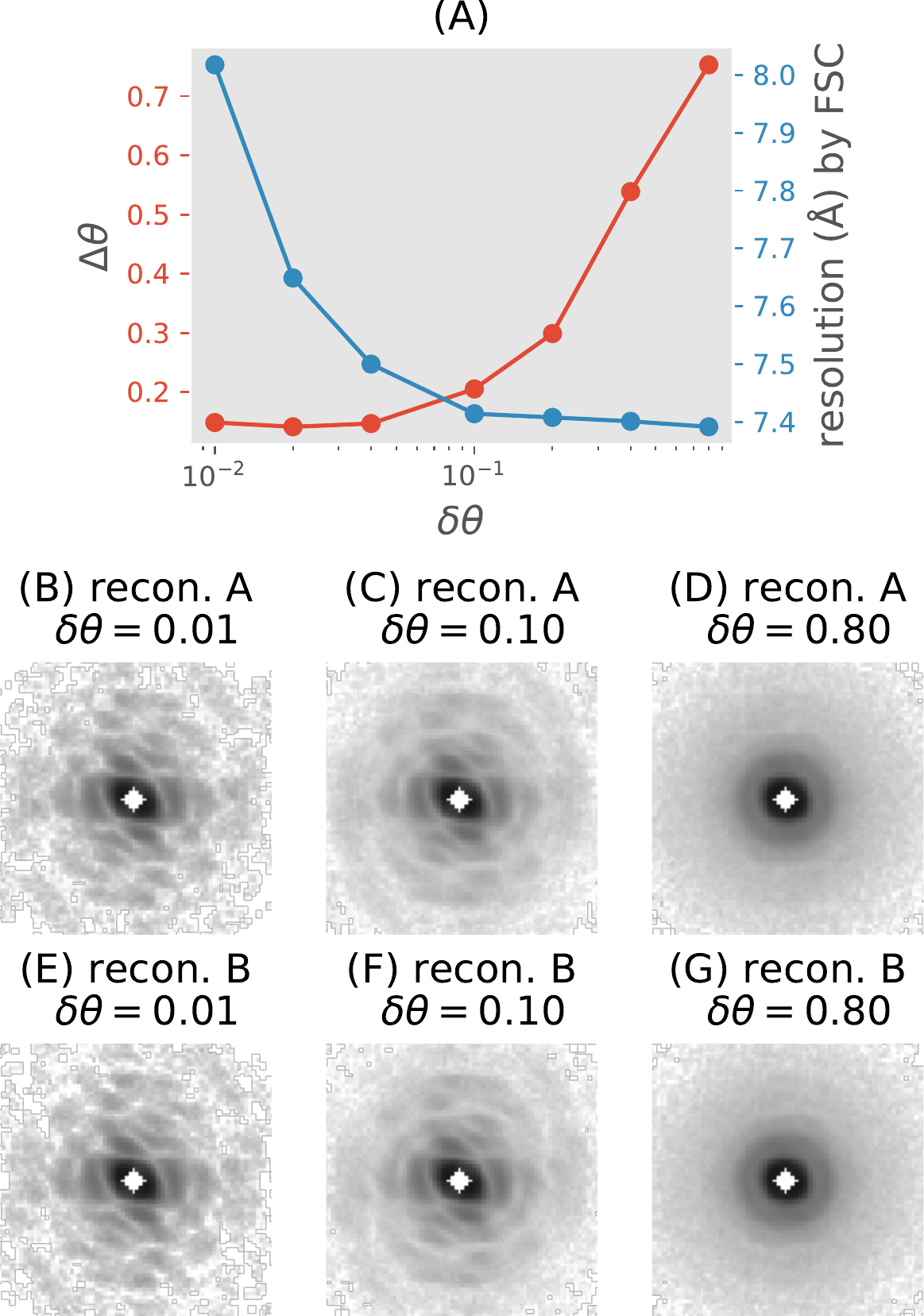}
    \caption{
    Fourier shell correlation (FSC) reports improved resolution despite increased orientational blurring. 
    Two disjoint SPI datasets were simulated, $A$ and $B$, each with 5000 patterns. 
    (A) The FSC was calculated for all pairs of reconstructions from the same dataset and with the same orientation blurring $\delta \theta$ (blue curve). 
    Diffraction volumes were reconstructed from each dataset by interpolating each pattern back into ten random orientations near the true one. The true variance of these orientations is denoted $\delta \theta^2$, which is proportional to the degree of deliberate orientation blurring.
    The orientation disconcurrence proposed in this paper, $\Delta \theta$ (red curve), was computed using a third smaller sentinel dataset (1000 patterns) not used in the reconstructions.
    For each dataset, seven 3D volumes were reconstructed by interpolating all patterns back into the 3D diffraction volume with $\delta \theta=\{0.01, 0.02, 0.04, 0.1, 0.2, 0.4, 0.8\}$. 
    (B-D) The central slices of one of the seven volumes for each $\delta \theta$ from dataset $A$, (E-G) and those from dataset $B$.
    }
    \label{fig:fsc_vs_dtheta}
\end{figure}

However, the use of FSC, as well as other proposed measures of reconstruction errors\cite{Yoon2016-mu, Liu2018-yu}, to characterize XFEL-SPI resolution suffers three main issues.
First, and most importantly, Fig.~\ref{fig:fsc_vs_dtheta} illustrates how the resolution reported using the popular half-bit FSC criterion actually improves with increased orientation blurring.
This occurs because XFEL-SPI reconstructions approach the same {\it virtual powder average} as their input patterns become more misoriented.
Consequently the `noise terms' between two independently reconstructed volumes (see Eqn.~(3) in \cite{Van_Heel2005-qk}) become correlated.
Hence the FSC measure, which is invariant to isotropic filtering, can paradoxically report better resolutions when the orientation uncertainty of patterns increases.
Second, the threshold criterion for determining resolution is controversial even in the cryo-electron microscopy community\cite{Van_Heel2005-qk,Liao2010-vl}.
This criterion is demonstrably dependent on the speckle sampling ratio (i.e.~size of realspace support), the symmetry of the particle, and assumes additive noise \cite{Van_Heel2005-qk}. 
Unfortunately, there are still prominent violations of these criteria \cite{Van_Heel2017-ot}. 
Third, to compute the FSC between two 3D volumes, their relative orientations must be accurately determined.

To circumvent some of these issues with FSC, we propose examining the source of correlations between two indpendently recontructed volumes: the `disconcurrence', inconsistency, and agreement between how these volumes orient individual patterns. 
A similar orientation-based approach to validation was explored by Tegze and Bortel \cite{Tegze2016-pi}, where they proposed using the fraction of patterns that are well-oriented to validate intensity reconstructions.
However, the so called $C$-factor that they proposed for validation only considered orientation precision but not accuracy or reproducibility. 
Hence, as that work suggested, the $C$-factor was susceptible to overfitting when too few patterns were used to reconstruct $W$.

It can be useful to recast the XFEL-SPI validation problem in information theoretic terms.
Indeed, information theory has been insightful for SPI \cite{Elser2009-ef} as well as coherent diffraction imaging \cite{Elser2011-ig, Jahn2017-oe}.
In fact, the half-bit criterion for FSC in cryo-electron microscopy\cite{Van_Heel2005-qk} established a connection between spatial resolution and information theory.  
There, however, the half-bit criterion merely referred to when the signal-to-noise ratio of an idealized noisy channel attained a value of $\sqrt{2}-1$. 
What this signal-to-noise ratio means for resolving spatial features within an object remains unclear. 

Looking farther back, Shannon's original proof of the noisy channel theorem was based on a straightforward encoding-decoding scheme \cite{Shannon1948-xn}. 
Below we show how Shannon's scheme can be explicitly constructed for the orientation determination problem in SPI.
Doing so, allows us to validate $W$ reconstructions using an orientation resolution that can be directly related to the mutual information of the SPI experiment.

\begin{figure}
    \centering
    \includegraphics[width=\textwidth]{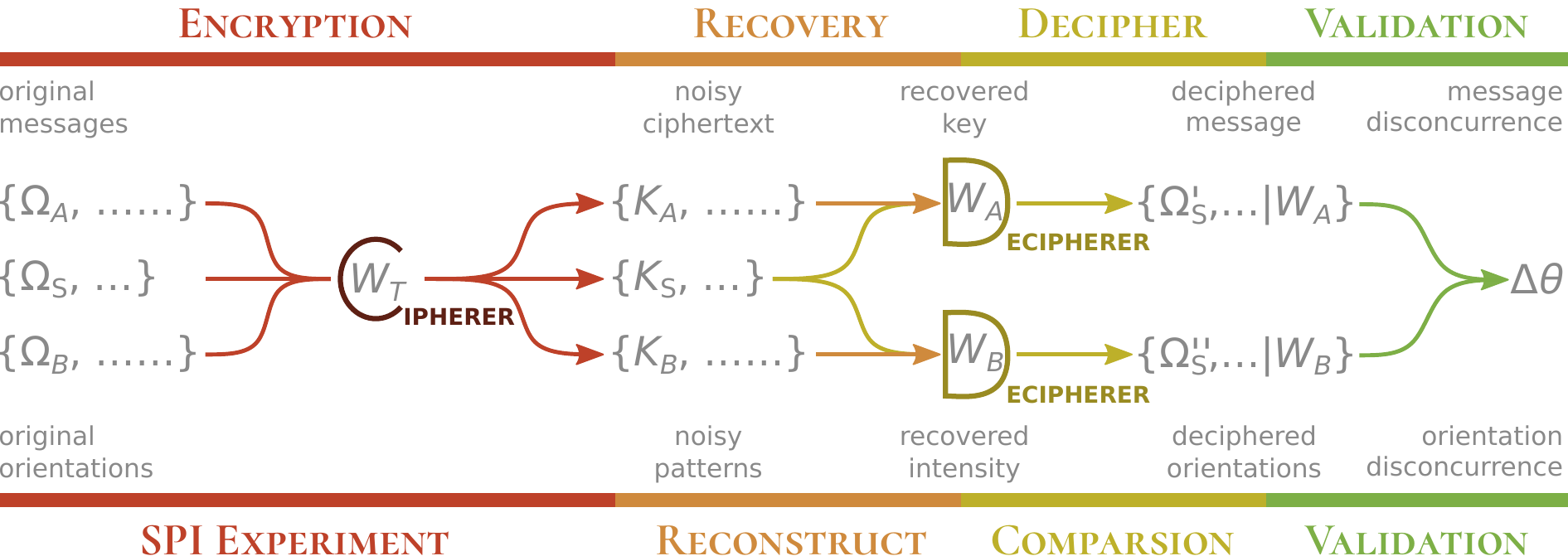}
    \caption{Analogy between `key-cracking' in cryptography (text in upper rows) and validation for single particle imaging (text in lower rows).}
    \label{fig:cryptography}
\end{figure}

An SPI reconstruction is similar to probabilistic symmetric-key cryptography, where \emph{plaintext messages} are encrypted into \emph{ciphertexts} using a \emph{correct key} plus a randomness scheme. 
Because of this randomness, the same plaintext message can produce different ciphertexts.

The analogous messages in an XFEL-SPI experiment are the hidden orientations of illuminated single particles \cite{lohCryptotomographyReconstructing3D2010}.
The experimental setup itself can be viewed as a cipher algorithm that encrypts these messages as noisy two-dimensional (2D) diffraction patterns. 
When these orientations (messages) are properly decrypted, the full three-dimensional (3D) diffraction volume of the target particle can be recovered.   
The conundrum for SPI, however, is that these orientations are best decrypted using the ground truth 3D diffraction volume.
Hence, reconstructing this diffraction volume can be viewed as `cracking' (i.e.~guessing) the correct symmetric key in probabilistic cryptography. 
Fig. \ref{fig:cryptography} shows the similarities between SPI-validation and key-cracking in cryptography, which has the following correspondence: 
\begin{itemize}
\item{} \emph{correct key} $\leftrightarrow$ ground truth 3D diffraction intensities; 
\item{} \emph{encryption cipher} $\leftrightarrow$ SPI experiment;
\item{} \emph{decryption cipher} $\leftrightarrow$ orientation inference scheme;
\item{} \emph{ciphertexts} $\leftrightarrow$ photon patterns collected in experiment; 
\item{} \emph{messages} $\leftrightarrow$ orientations of individual photon patterns.
\end{itemize} 
Algorithms that discover the orientations of SPI patterns \cite{Loh2009-eg, Bortel2011-za, Kassemeyer2013-yx, Tegze2013-cx}, analogously, try to recover the unknown key (i.e.~3D diffraction intensities) given many ciphertexts (i.e.~photon patterns).

Now let us consider how one can check/validate the accuracy/correctness of a recovered key, absent the ground truth. 
An obvious method is to determine whether the recovered key is consistent with known prior constraints or independent measurements. 
Such external validations, however, are not always possible in SPI especially when resolving novel structural forms. 

We know that a correct key must decipher each ciphertext into a unique message. 
However, this \emph{uniqueness alone is insufficient to determine correctness}, since wrong keys given to a deterministic cipher can yield unique but wrong decipherments. 
An example of this occurs when a recovered key overfits to a set of ciphertexts.  
Nevertheless, we can exploit this uniqueness requirement to design a scheme that detects if at least one of two candidate keys is incorrect. 

Suppose we are given two disjoint sets of ciphertexts ($\{K_A\}, \{K_B\}$) that are encrypted by the same solution key $W_T$. 
We can independently recover two keys ($W_A, W_B$), one from each set of ciphertexts. 
Disagreements between how these two keys decipher a third hidden set of ciphertexts $\{K_\text{S}\}$ betrays the incorrectness of at least one of these two keys. 
If the first two sets of ciphertexts are sufficiently large and randomly chosen then both candidate keys are likely incorrect.

Owing to the randomness in probabilistic encryption, it is practically impossible to guarantee a perfectly accurate key given only a finite number of noisy ciphertexts. 
Analogously, we cannot perfectly recover the ground truth SPI diffraction volume only from a finite number of noisy, incomplete photon patterns. 
Consequently, any pair of recovered keys must differ measurably from each other. 
This difference quantifies the {\it decryption precision} of these keys, which is the lower bound of their {\it decryption accuracies}. 

Back to the SPI data analysis, we wish to find the difference in how two independently reconstructed volumes $W_A$ and $W_B$ decrypt the orientations of a third disjoint set of {\it sentinel photon patterns}, $\{K_\text{S}\}$.
This difference in decryption increases if the disagreement between $W_A$ and $W_B$ increases.
More importantly, it also increases as either volume departs farther from the hidden ground truth volume $W_T$. 
We refer to this difference as the {\it orientation disconcurrence} between these two volumes. 
The procedure to compute this disconcurrence is outlined below (see Fig. \ref{fig:cryptography}).

\begin{figure}
    \centering
    \includegraphics[width=0.7\textwidth]{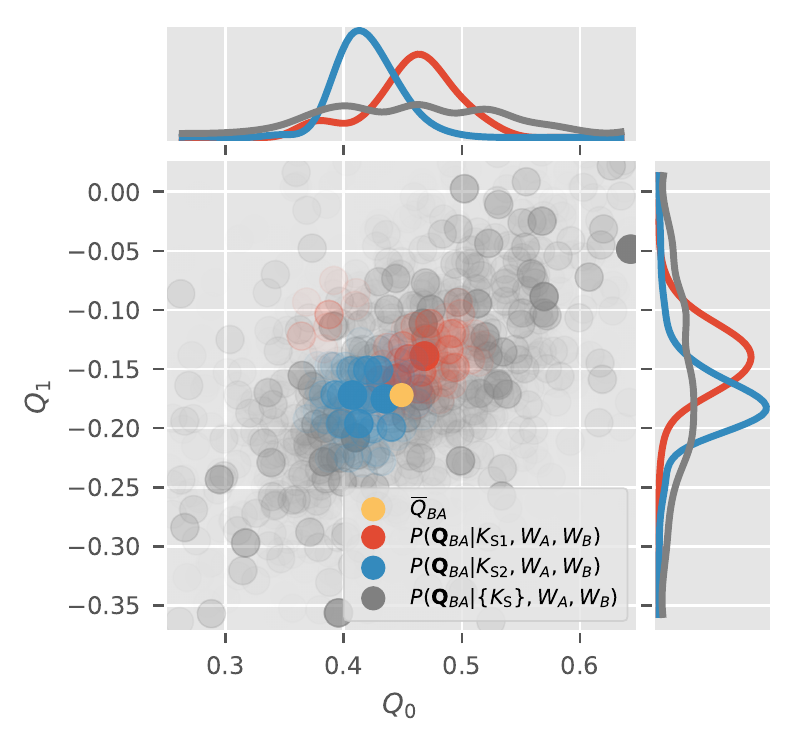}
    \caption{Clustering of the angular displacement distribution (ADD) for $1000$ sentinel patterns given two independently reconstructed volumes $W_A$ and $W_B$, in the space of possible unit quaternions. Only the first two components of these quaternions ($Q_0, Q_1$) are shown. The disks represent the set of most significant relative quaternions given each sentinel pattern, $\{\bbQ_{BA} \given K_\text{S}\}$, as defined by all possible pairs of those in \myeqref{eqn:imptQuat}. The opacities of these disks are proportional to the value of the ADD at these quaternions. The blue and red disks represent the ADDs for two specific sentinel patterns respectively. The yellow disk shows the average overall rotation $\overline{Q}_{BA}$ as defined in \myeqref{eqn:OverallOrientation}.
    }
    \label{fig:4dcluster}
\end{figure}

\begin{enumerate}
\item{} Partition the XFEL-SPI photon patterns $\{K\}$ into three disjoint sets: two larger and equally sized sets, $\{K_A\}$ and $\{K_B\}$, for reconstructions; and a third, smaller set of unseen {\it sentinel patterns} $\{K_\text{S}\}$ to measure orientation disconcurrence.
\item{} Using any algorithm you desire, reconstruct two 3D intensities from the two larger sets of patterns: $\{K_A\} \to W_A$, and $\{K_B\} \to W_B$.
\item{} For each sentinel pattern $K_\text{S}$, compute the orientation posterior distribution (OPD, defined in \myeqref{eqn:posterior-likelihood}) of the reconstructed volumes $W_A$ and $W_B$. This is the probability that $K_\text{S}$ corresponds to the Ewald sphere section of orientation $\Omega$ in each reconstructed volume (i.e.~$P(\Omega_A|K_\text{S}, W_A)$ and $P(\Omega_B|K_\text{S}, W_B)$). This step creates $2\,|\{K_\text{S}\}|$ distributions, two for each sentinel pattern, where $|\{K_\text{S}\}|$ is the number of sentinel patterns used.
\item{} Next, we compute the angular displacement distribution (ADD, defined in \myeqref{eqn:ADD-single}) of the sentinel patterns from the OPD of $W_A$ and $W_B$. The ADD for each sentinel pattern $K_\text{S}$ (the red or blue distribution in Fig.~\ref{fig:4dcluster}) is essentially a convolution of OPD$_A$ and OPD$_B$ over the space of relative orientations between $W_A$ and $W_B$. If OPD$_A$ and OPD$_B$ were delta functions, then this convolution peaks at the relative orientation between $W_A$ and $W_B$. The ADD$_{AB}$ (the grey distribution in Fig.~\ref{fig:4dcluster}), which is the normalized sum of these convolutions for all sentinel patterns (\myeqref{eqn:ADD-all}), is the distribution of relative orientations between $W_A$ and $W_B$ as `measured by' $\{K_\text{S}\}$.
\item{} Finally, from the ADD of all the sentinel patterns between the volumes $W_A$ and $W_B$, estimate their orientation disconcurrence.
\end{enumerate}

\section{Results}
\subsection{Measures of orientation uncertainties.}\label{ssec:typesOfUncertainty}

\begin{figure}
    \includegraphics[width=\textwidth]{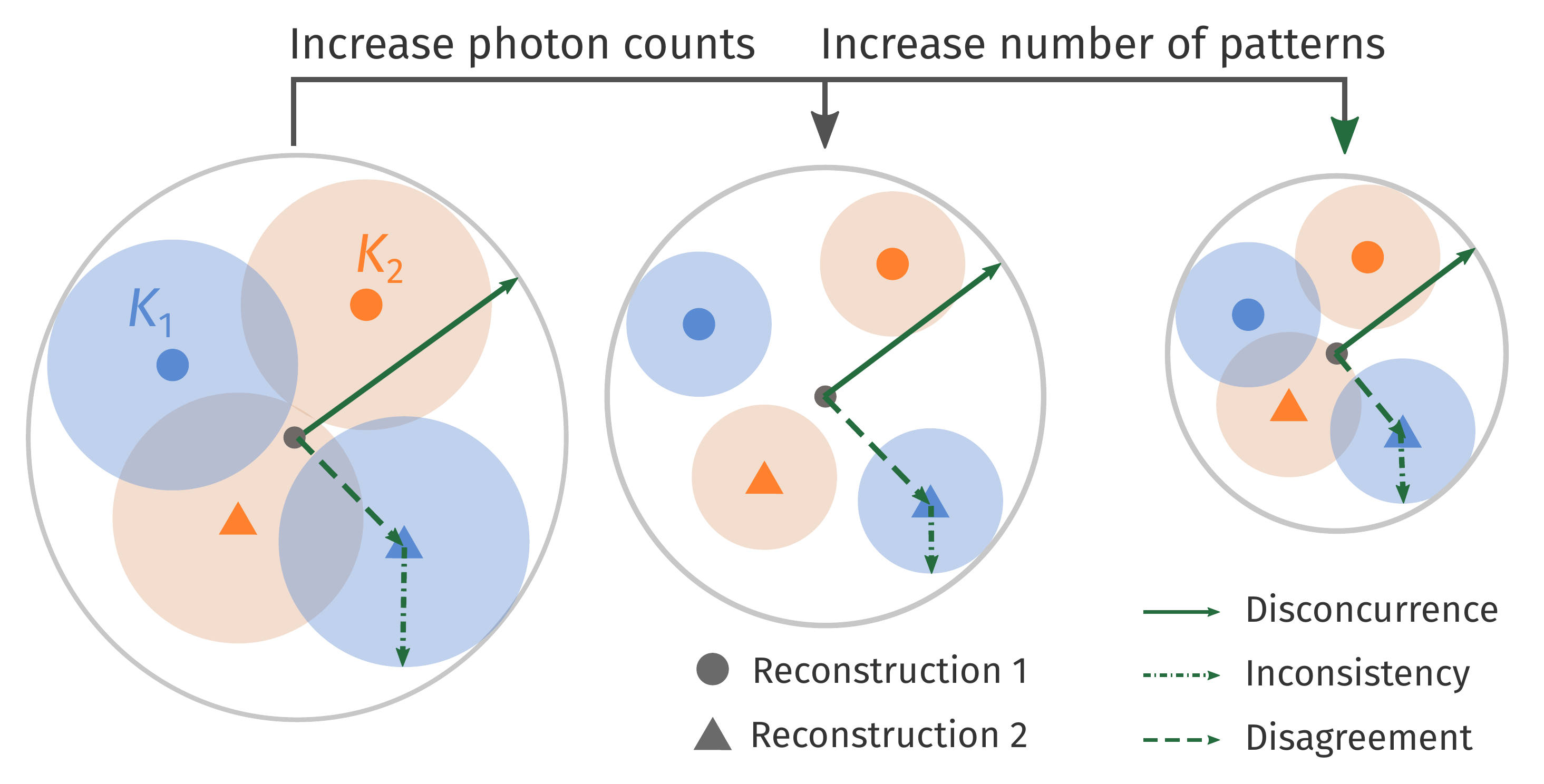}
    \caption{The orientation disconcurrence for two sentinel patterns ($K_1$ in blue, and $K_2$ in orange) consists of two parts: the inconsistency that each model orients sentinel patterns (disk spanned by dashed-dotted radii), and the disagreement between how different models orient these patterns (disk spanned by dashed radii). These aspects are affected by the photon counts per pattern ($N$) and the number of patterns ($\Mdata$) respectively.}
    \label{fig:consis_argee}
\end{figure}

The orientation \emph{disconcurrence} between two independently reconstructed volumes comprises two aspects: \emph{inconsistency} and \emph{disagreement}. 
By the cryptographic analogy, the first aspect characterizes how \emph{consistently} each volume separately decrypts the orientations of sentinel patterns; the second aspect describes how often the decryptions of two (or more) volumes \emph{mutually agree}. 
These concepts are illustrated in Fig. \ref{fig:consis_argee}, and defined below.

In the following numerical simulations, we use the disconcurrence between independent reconstructions from the same scatterer to estimate the lower bound of their correctness. 
Recall that this procedure requires partitioning a set of photon patterns into three disjoint sets ($\{K_A\}, \{K_B\}, \{K_\text{S}\}$). 
We reconstruct two 3D intensities from the first two sets ($W_A$ and $W_B$ respectively), while the last sentinel set is reserved for validation. 
Unlike an actual experiment, the true solution intensities $W_T$ that generated these patterns are known in these simulations, and will provide useful insights.
Given these definitions, let us consider different orientation measures at the end of the procedure outlined at the end introduction section.  

\begin{enumerate}
    \item Measure of orientation disconcurrence:
        $\Delta \theta_\text{c}(W_A, W_B) $ (\myeqref{eqn:OrientationConcurrence})
    is computed from the width of the angular displacement distribution (ADD) between intensities $W_A$ and $W_B$ that are independently reconstructed from two disjoint sets of patterns. $\Delta \theta_\text{c}$ measures the difference between the orientations of specific sentinel patterns within $W_A$ and $W_B$, despite having aligned the centroids of these two distributions (i.e.~overall orientations of $W_A$ and $W_B$).
    
    \item 
    Measure of average orientation inconsistency:
    \begin{equation}
        \Delta \theta_\text{i}(W_A, W_B) =
        \sqrt{\frac{1}{2}
        \sum_{i\in \{A,B\}}
    \Delta\theta^2_\text{c}(W_i, W_i)}\;.
    \label{eqn:OrientationConsistency}
    \end{equation}
    This is the root-mean-squared (RMS) angular width of the autocorrelation of $W_A$'s and $W_B$'s orientation posterior distribution (OPD), which is equivalent to repeating the intensity model labels in \myeqref{eqn:Theta}.
    In Fig.~\ref{fig:4dcluster}, the angular width of the blue and red points show the orientation inconsistency for decryption the orientations of two sentinel patterns ($K_1$ and $K_2$). 
    The RMS of $\Delta\theta^2_\text{c}(W_A, W_A)$ and  $\Delta\theta^2(W_B, W_B)$ is used to approximate the angular width (red or blue distribution) in Fig.~\ref{fig:4dcluster},
    because it is expensive to calculate the inconsistency between $W_A$ and $W_B$ for each sentinel patterns and it is a good approximation when the OPD is assumed to be a Gaussian distribution (see more details in
    Section~\ref{appendix:1dmodel}). 
    Thus $\Delta \theta_\text{i}$ simply averages this width over all sentinel patterns and both reconstructions $W_A$ and $W_B$.  
    
    \item Measure of orientation disagreement: 
    \begin{align}
        &\Delta \theta_\text{a} (W_A, W_B) = \sqrt{\left(\Delta \theta_\text{c} (W_A, W_B)\right)^2 - \left(\Delta \theta_\text{i}(W_A, W_B)\right)^2}\; ,
        \label{eqn:OrientationAgreement}
    \end{align} which is the angular displacement between reconstructions $W_A$ and $W_B$ that is not due to an overall rotation between the two volumes, nor from the angular width $\Delta \theta_\text{i}$ of the OPD.
    In Section~\ref{appendix:1dmodel}, this relation is illustrated with a 1D model in more detail.
    
    \item Measure of orientation inconsistency given the ground truth: 
        \begin{equation}
            \Delta \theta_\text{i}^{\ast}=\Delta \theta_\text{c}(W_T, W_T)\; ,
        \label{eqn:OrientationConsistencyGT}
        \end{equation}
    which measures the angular width of the OPD in determining the patterns' orientations given the ground truth $W_T$. 
    With enough patterns in $\{K_A\}$ and $\{K_B\}$, such that $W_A$ and $W_B$ do not over-fit to their respective photon patterns, we expect $\Delta \theta_\text{i} \geq \Delta \theta_\text{i}^\ast$. 
    
    \item Measure of orientation disconcurrence with ground truth: 
    \begin{equation}
        \Delta \theta^{\ast}_\text{c}(W_A)=\Delta \theta_\text{c}(W_A, W_T) \; ,
        \label{eqn:OrientationConcurrenceGT}
    \end{equation} which is the angular width of the ADD between the reconstructed and ground truth intensity volumes ($W_A$ vs $W_T$ respectively). Notice that $\Delta \theta_\text{c}$ is identical to $\Delta \theta^{\ast}_\text{c}$ above if we replaced $W_B \to W_T$. Hence, {\it $\Delta \theta^{\ast}_\text{c}$ is essentially the orientation disconcurrence between $W_A$ and the ground truth.}
    
    \item Measure of average orientation disconcurrence with ground truth:
    \begin{equation}
        \langle \Delta \theta^{\ast}_\text{c} \rangle=\sqrt{\frac{1}{2} \left[ \left(\Delta \theta^{\ast}_\text{c} (W_A)\right)^2 + \left(\Delta \theta^{\ast}_\text{c} (W_B)\right)^2 \right] } \; ,
    \label{eqn:AverageOrientationConcurrenceGT}
    \end{equation} which is the average angular width of the ADDs between the reconstructed versus the ground truth intensity volumes ($W_A, W_B$ vs $W_T$ respectively).
    {\it If only two volumes were reconstructed, $W_A$ and $W_B$, then $\langle \Delta \theta^{\ast}_{c} \rangle$ represents the average orientation disconcurrence against the ground truth}.
\end{enumerate}

\subsection{Factors that influence disconcurrence.}\label{ssec:results}
Many experimental factors influence the orientation disconcurrence of an SPI intensity reconstruction including: incident photon fluence, number of photon patterns from single particles, resolution and sampling of each pattern, amount of missing detector data (i.e.~beamstop, gaps in compound detectors, inactive pixels), extent of photon background (i.e.~from particles' incoherent scattering or stray light sources), degree of structural heterogeneity between particles in the ensemble. 
The choice of algorithms and their parameters used to reconstruct the intensities also play important roles. 
Furthermore, the symmetries of the scatterer itself can also affect how the ADD is intepreted (see Fig. \ref{fig:oct_cluster} and Methods). 

In this section, we focus on three of these factors: the average number of photons per pattern $N$, the fineness of orientation space sampling by reconstruction algorithms, and the number of patterns $\Mdata$.
In each scenario studied below, we simulated diffraction patterns with a small 105 kDa protein (PDB code, 4ZW6 \cite{drinkwater2016potent}) under experimental conditions that were modeled after those at the Tender X-ray endstation at the Linac Coherent Light Source (see Table \ref{table:DragonflyParameters}). 
We then used the EMC algorithm to reconstruct two independent 3D volumes each from disjoint sets $\{K_A\}, \{K_B\}$, each with $\Mdata$ patterns. 
For each test condition, a single set of 1000 sentinel patterns was reserved $\{K_\text{S}\}$ to evaluate the six types of $\Delta \theta$ listed above. 

\begin{table}
\caption{Range of parameters used to simulate XFEL-SPI photon patterns in this paper.}
\label{table:DragonflyParameters}
\begin{tabular}{ll}
\hline
parameter & value  \\
\hline
photon wavelength (\si{\angstrom})  &  \num{3.4}\\
detector distance (\si{mm})  &  \num{300}\\
detector pixel size (\si{mm}) &  \num{1.2} \\
detector size (\si{pixel})&  $\num{100}\times \num{100}$ \\
beamstop radius (\si{pixel})& \num{10}\\
photon fluence (\si{photons\cdot \micro m^{-2}}) & \numrange{1e13}{5e13}\\
focal area (\si{\micro\meter^2})\tnote{\dag} & 
\numrange[parse-numbers = false]{0.33^2}{0.15^2}\\
\hline
\end{tabular}
 \begin{tablenotes}
  \item[*] $\dag$ Assume: incident beam energy \SI{3}{mJ}, transmission efficiency $20\%$.
  \item[*] $\ddagger$ A binned detector is used here for computational efficiency. 
  \end{tablenotes}
\end{table}

\begin{figure}
    \includegraphics[width=0.6\textwidth]{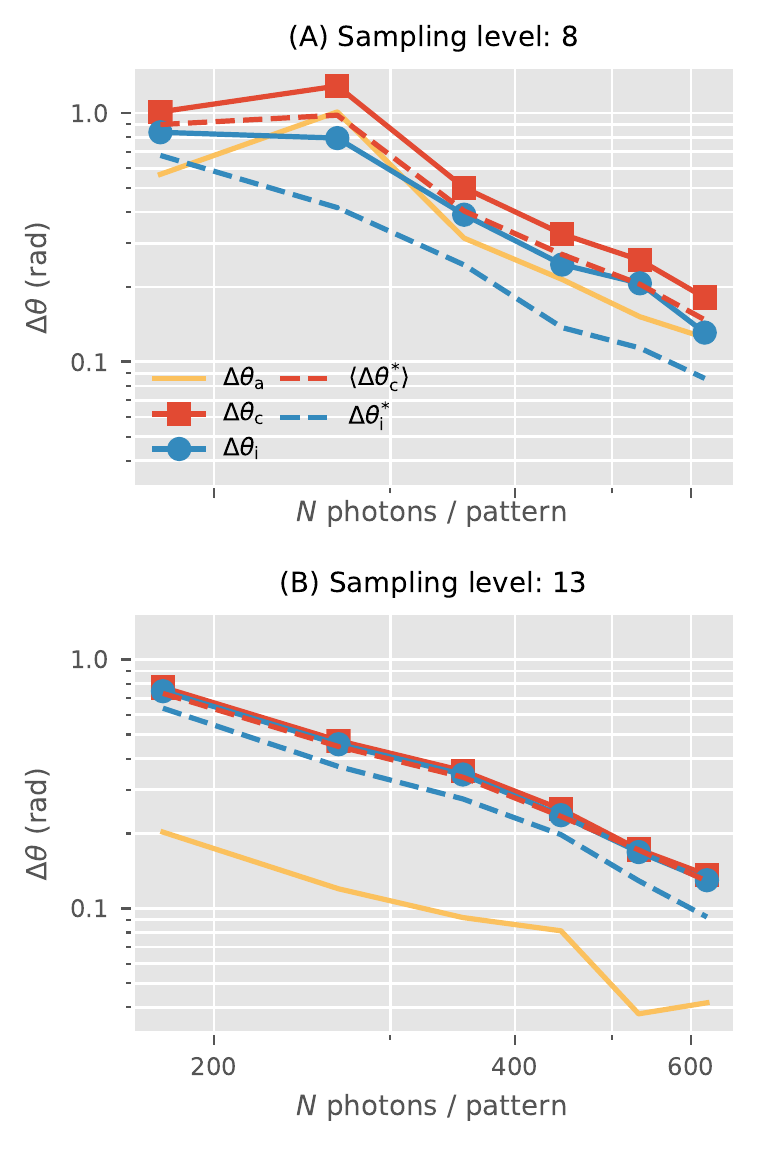}
    \caption{
        Effects of incident photon counts per pattern and sampling fineness of the latent orientation space. 
        Each data point compares two 3D intensity reconstructions with $5000$ photon patterns (solid lines), or each one of them with a ground truth 3D intensity volume (dashed lines). The rotation
        group is sampled with refinement levels $n=8$ or $n=13$. 
        As the average photon counts per pattern increases, all varieties of angular uncertainties specified in Section \ref{ssec:typesOfUncertainty} decrease. The uncertainties involving the ground truth ($*$-superscript, dashed lines here) are typically lower than those with only the reconstructed volumes (solid lines). 
        Finer orientation sampling reduces all orientation uncertainties. Furthermore, orientation disconcurrence ($\Delta \theta_\text{c}$, red) is dominated by inconsistency ($\Delta \theta_\text{i}$, blue) as orientation disagreement ($\Delta \theta_\text{a}$, yellow) is suppressed.    
    } 
    \label{fig:dtheta-photon}
\end{figure}


The average number of photons per diffraction pattern ($N$) is directly related to the mutual information for inferring latent parameters (e.g.~orientations) as well as the particle's structure \cite{Loh2009-eg}. 
$N$ depends on the brightness of the x-ray beam, the size of the x-ray focus (i.e.~beam intensity), as well as the relative alignment between particle and x-ray beams. 
In general, all six types of $\Delta\theta$ fall when $N$ increases in Fig.~\ref{fig:dtheta-photon}. 
Simply put, more photons per pattern reduces orientation disagreement and inconsistency, hence disconcurrence. 
Additionally, the orientation disconcurrence between $W_A$ and $W_B$ falls with their respective
disconcurrences with the ground truth $W_T$. 
This correspondences is consistent with the fact that uniqueness is a necessary condition for correctness (i.e.~`precision $\leq$ accuracy').


How finely orientations are sampled in XFEL-SPI reconstruction algorithms impacts the quality of reconstructed results \cite{Loh2009-eg}.
Recall, this sampling fineness is different from the adaptive refinement scheme for OPD and ADD \myeqref{eqn:imptQuat}: the former pertains to the reconstruction algorithm, while the latter evaluates the reconstructed results.  
Fig.~\ref{fig:dtheta-photon} shows that a higher sampling level in the EMC reconstruction algorithm generally reduces all alignment uncertainties $\Delta\theta$. 
While the various forms of $\Delta \theta$ have a noticeable spread at $n=8$ orientation sampling, this spread significantly reduces when this sampling fineness is increased to $n=13$. 
Numerically, we found the average angular separation between the quasi-uniform unit quaternions samples to be 0.161 and 0.099 radians respectively. 
This figure complements the information-theoretic heuristic for deciding sampling sufficiency in \cite{Loh2009-eg}. 
With sufficient sampling, Fig.~\ref{fig:dtheta-photon} shows that the orientation disconcurrence is dominated by the orientation inconsistency rather than orientation disagreement: $\Delta \theta_\text{c} (W_A, W_B) \approx \Delta \theta_\text{i} (W_A, W_B) > \Delta \theta_\text{a} (W_A, W_B)$.

\begin{figure}
    \includegraphics[width=\textwidth]{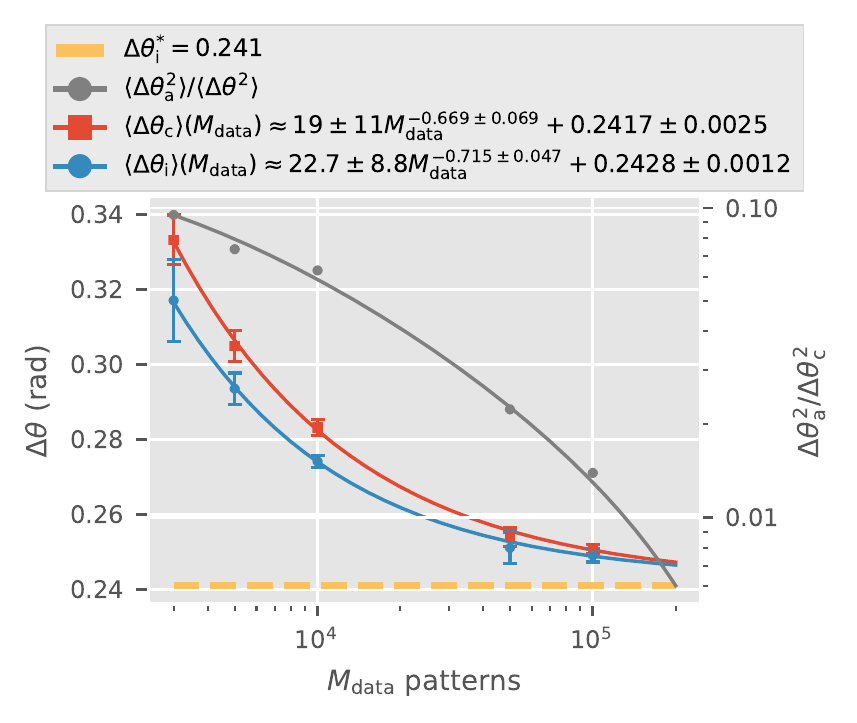}
    \caption{Orientation disconcurrence ($\Delta \theta_\text{c}$) and inconsistency ($\Delta \theta_\text{i}$) converge to $\Delta \theta_\text{i}^*$ as the number of patterns ($\Mdata$) increase. Each dot and its error bars represent the average and standard deviation of $\Delta \theta$ of all pairs among five reconstructions from four different disjoint datasets (average of 355 photons/pattern, rotation group sampling $n=13$). The same 1000 sentinel patterns are used in all four instances. The ratio of orientation disagreement $\Delta\theta_\text{a}$ to disconcurrence $\Delta\theta_\text{c}$, which is represented by the grey curve (labeled on right vertical axis), decreases with increasing $\Mdata$.
      }
    \label{fig:dtheta-pattern}
\end{figure}

In an SPI experiment the number of SPI patterns, $\Mdata$, is a product of the fraction of particles that are illuminated by x-ray pulses (i.e.~hit-rate), the pulse repetition rate, and the total experiment time. 
One intuitively expects that reconstructions improve with larger $\Mdata$, which Fig.~\ref{fig:dtheta-pattern} confirms. 
The intrinsic orientation inconsistency of each reconstruction, $\Delta\theta_\text{i}$, falls with more patterns (blue curve). 
The orientation disconcurrence $\Delta\theta_\text{c}$, likewise, also falls with more patterns. 

 We found that in Fig.~\ref{fig:dtheta-pattern} that $\Delta \theta_\text{c}$ and $\Delta \theta_\text{i}$ both decrease numerically with the number of patterns as $\alpha \, \Mdata^{-\beta} + \Delta\theta_\text{i}^{\ast}$, where $\alpha$ is a multiplicative constant, $\beta$ is a real positive number, and $\Delta\theta_\text{i}^{\ast}$ is the angular width of the OPD given the patterns $\{K_\text{S}\}$ and ground truth model.
Although $\Delta \theta_\text{c} \to \Delta \theta_\text{i}^{\ast}$ as $\Mdata \to \infty$, we can only assert that the reconstructed pairs of models ($W_A$ and $W_B$) are closer to each other, but not whether either are close to the ground truth $W_T$.
The former is evident from the ratio of orientation disagreement against disconcurrence, $\Delta\theta_\text{a}^2 / \Delta\theta_\text{c}^2$ (gray dots in Fig.~\ref{fig:dtheta-pattern}): increasing $\Mdata$ eliminates orientation disagreements ($\Delta \theta_a$) between two independent reconstructions faster than intrinsic inconsistency ($\Delta \theta_\text{i}$). 
Using \myeqref{eqn:OrientationAgreement} and the fitted forms in Fig.~\ref{fig:dtheta-pattern}, this vanishing of the orientation disagreement becomes clear: 
\begin{align}
        \Delta\theta_\text{a} &= \sqrt{\Delta\theta_\text{c}^2 - \Delta\theta_\text{i}^2} \nonumber \\
        &=\sqrt{\big(\alpha_\text{c} \Mdata^{-\beta_\text{c}} + \gamma_\text{c}\big)^2-
            \big(\alpha_\text{i}\Mdata^{-\beta_\text{i}} + \gamma_\text{i}\big)^2 }\;  \nonumber \\
       &\approx  \Mdata^{-\beta_\text{c}/2} \sqrt{\left(\alpha_\text{c} + 2 \gamma \right)\alpha_\text{c}}\; ,
       \label{eq:power_law_sigmaa}
\end{align} 
where we assumed $\beta_\text{c} < \beta_\text{i}$, and $\gamma_\text{c} \approx \gamma_\text{i}=\gamma$.
Obviously, when $\Mdata$ approaches infinity, $\Delta\theta_\text{a}$ gets close to $0$. 
Simply put, as $\Mdata$ increases independently reconstructed volumes become more unique but not necessarily more correct. 

\subsection{Relating $\Delta \theta$ to spatial resolution.}\label{subsec:discuss_resolution}
\label{subsec:dtheta-resolution}
The 3D speckles in the reconstructed diffraction volume whose angular width are smaller or comparable to $\Delta \theta_\text{c}$ will lose contrast, hence spatial resolution. 
Let us denote the full angular width of these 3D speckles as $2\Delta \theta_{\text{sp}} (\bbqRes)$ at spatial resolution $\bbqRes$. 
Naturally, the reconstructions become orientation-limited at the resolution where $\Delta \theta_{\text{sp}} (\bbqRes)$ approaches the width of OPD which is about $\Delta\theta_\text{c}/\sqrt{2}$ (Section~\ref{appendix:1dmodel}).

\begin{figure}
    \centering
    \includegraphics[width=0.8\textwidth]{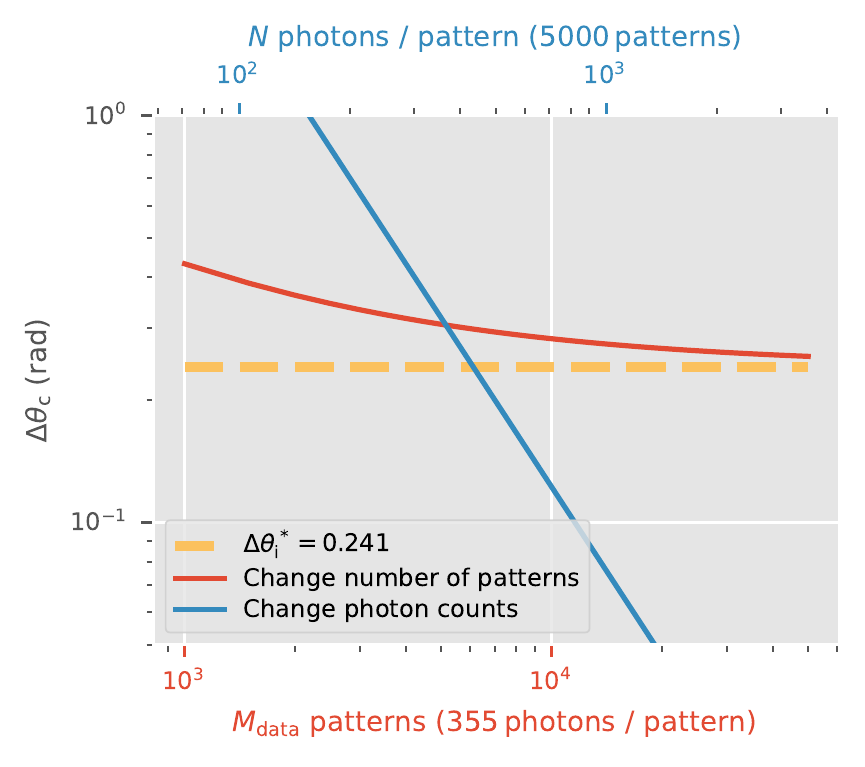}
    \caption{This figure shows how $\Delta\theta_\text{c}$
    changes by increasing number of patterns (red curve, with $N\approx 355$) or number of photons per pattern (blue curve, with $\Mdata=5000$). The measure of orientation inconsistency given the ground truth, $\Delta \theta_\text{i}^\ast$ (yellow), is computed for $N\approx 355$.}
    \label{fig:dtheta_pho-pat}
\end{figure}

Fig.~\ref{fig:dtheta_pho-pat} shows that it is possible for reconstructions whose orientation disconcurrence is smaller than the angular width of a single pixel at the edge of the detector $\Delta \theta_{\text{pix}}$.
This situation occurs with very high average number of photons per pattern ($N \gg 1$), abundant patterns ($\Mdata \gg 1$), and sufficiently fine sampling of the rotation group during reconstructions (Fig.~\ref{fig:dtheta-photon}). 
Thus, the dynamic range and contrast of the reconstructed 3D diffraction speckles are high up to the detector's maximum captured resolution ($\bbqRes_\text{max}$), which allows us to distinguish arbitrarily small angular variations between actual diffraction patterns. 

We must remember that the reconstructed diffraction volume $W$ does not explicitly contain spatial information beyond the maximum spatial resolution $\bbqRes_\text{max}$. 
So even if $\Delta \theta_\text{c} \ll \Delta \theta_\text{pix}$, we can only say that spatial resolution is not orientation limited. 
Perhaps with additional priors about the structure of the particle (e.g.~know sequence, similar structure known, atomicity, etc) is might be possible to extend the resolution beyond $\bbqRes_{\text{max}}$. 
But such extensions are beyond the scope of this discussion.

It should now be clear that orientation disconcurrence relates to how effectively one can resolve the orientation of an average SPI photon pattern.
From this section, it should also be clear that spatial resolution can be limited by large orientation disconcurrences.
However, it is premature to define spatial resolution only in terms of orientation concurrence, especially since a decryption scheme for the spatial resolution (similar to Fig. \ref{fig:cryptography}) is absent.

\subsection{Data sufficiency and mutual information.}\label{subsec:DiscussDataSufficiency}
The question `how many patterns are sufficient?' frequently occur in an XFEL-SPI experiment.
The answer to this hypothetical question determines if a proposed experiment is `feasible', as well as how many different samples to inject during the precious dozens of hours of XFEL beamtime allocated to each user group.
Orientation disconcurrence can be used to define data sufficiency: when the number of patterns gives a disconcurrence smaller than the angular width of speckles at a target resolution $\text{\bf q}_\text{target}$: 
\begin{equation}
    2\cdot\frac{\Delta \theta_\text{c}}{\sqrt{2}} \leq \theta_\text{sp} (\bbqRes_\text{target}) \; . \label{eqn:resolutionCondition} 
\end{equation}
If the ADD peak in Fig.~\ref{fig:4dcluster} were compact and locally Gaussian (Section~\ref{appendix:1dmodel}), this last condition means that approximately $74\%$ ($2\sigma$ criterion) of the oriented sentinel patterns should intersect their target 3D speckle at resolution $\bbqRes_\text{target}$.   

With the disconcurrence target defined, we can extrapolate data sufficiency with bootstrapping. 
Given $\Mdata$ total patterns, one can compute $\Delta\theta_\text{c} (\Mdata)$ for pairs of models reconstructed from random, non-overlapping, equal subsets from the full $\Mdata$ dataset similar to the data points in Fig.~\ref{fig:dtheta-pattern}. 
Repeating this procedure via a simple bootstrapping scheme gives the orientation disconcurrence curves in Fig.~\ref{fig:dtheta-pattern}.
These curves fit reasonably well to a lifted exponential, $\Delta\theta_\text{c} = \alpha_\text{c} \Mdata^{- \beta_\text{c}} + \gamma_\text{c}$. 
The shrinking error bars on $\Delta \theta_\text{c}$ from bootstrapping with increasing $\Mdata$ in Fig.~\ref{fig:dtheta-pattern} suggests that this fit requires sufficiently many patterns to be robust.

Using only $\Mdata$ experimentally measured photon patterns, the lifted exponential fit allows us to extrapolate data sufficiency, as defined by orientation disconcurrence, to at least two different scenarios. 
First, if $\Delta \theta_\text{c}(\Mdata \leq M/2)$ were computed between pairs of reconstructed volumes each using up to $M/2$ bootstrapped photon patterns, then the angular uncertainty of a single volume with all $M$ patterns can be extrapolated using the fit: $\Delta \theta_\text{c}(\Mdata=M) = \alpha_\text{c} M^{\beta_\text{c}} + \gamma_\text{c}$.
A similar extrapolation from bootstrapped reconstructions was proposed to define spatial resolution in cryo-electron microscopy\cite{rosenthalOptimalDeterminationParticle2003}.

Should the target orientation disconcurrence be the angular width of a single pixel at the edge of the detector, $\Delta\theta_\text{c}=\Delta\theta_{\text{pix}}(\bbqRes_\text{max})$, then $\gamma_\text{c} < \Delta\theta_\text{pix}(\bbqRes_\text{max})$ is required. 
If this requirement is satisfied, then $\frac{1}{\beta_\text{c}}\log{\left[\alpha_\text{c}/(\Delta\theta_{\text{pix}}(\bbqRes_\text{max}) - \gamma_\text{c}) \right]}$ patterns are needed to reach this target.

The lifted power law form of $\Delta\theta_\text{c} = \alpha_\text{c} \Mdata^{- \beta_\text{c}} + \gamma_\text{c}$ in Fig.~\ref{fig:dtheta-pattern} allows us to parametrize data sufficiency in an information-theoretic sense. 
Essentially, the mutual information here can be defined as the reduction in the entropy of orienting an average sentinel pattern give a set of $\Mdata$ photon patterns $\{K\}$.  
Ignoring factors of order unity, this mutual information, is approximately
\begin{align}
    I(\Omega_{\text{S}}, \{K\}) &\approx \log\left(\frac{2 \pi^2}{\Delta \theta_\text{c}^3} \right) \nonumber \\
    &\approx \log \left( \frac{2 \pi^2}{ \Delta \theta_\text{i}^{\ast 3}} \right) - \frac{3\alpha_\text{c}}{\Delta \theta_\text{i}^\ast} \Mdata^{-\beta_\text{c}} \; , \label{eqn:MutualInformationDecoding}
\end{align}
assuming $\Mdata \gg 1$. 

\myeqref{eqn:MutualInformationDecoding} contains two intuitive results.
First, this mutual information is bounded from above by that when the solution intensities are known: $\log \left( 2 \pi^2 / (\Delta \theta_\text{i}^\ast)^3 \right)$. 
This upper bound can be viewed as the SPI channel capacity for decryption orientations, and is computed in the same manner as the mutual information $I(K,\Omega)|_W$ in \cite{Loh2009-eg}.
Second, the mutual information for decryption orientations increases with the number of patterns.
This assumes that $\alpha_\text{c}/\Delta \theta_\text{i}^\ast > 0$ and $\beta_\text{c} > 0$, which are manifest in Fig.~\ref{fig:dtheta-pattern}. 
 Furthermore, $\beta_\text{c} > 0.5$ in Fig.~\ref{fig:dtheta-pattern}, which is better than one would expect if patterns were mutually independent (i.e.~$\beta_\text{c} = 0$).  
 This `co-dependence' arises because additional patterns can improve the reconstructed volumes, which in turn help earlier patterns distribute their photons more precisely into orientation classes. 
 
 \subsection{Focal spot size affects hit rate and orientation disconcurrence.}
The linear size of the XFEL focus $L_\text{focus}$ is a critical parameter in an SPI experiment (see Table \ref{table:DragonflyParameters}). 
This choice of focus size can be paraphrased simply: given a fixed total number of photons per XFEL pulse, would it be better to `distribute' them into more patterns with fewer photons each, or fewer patterns with more photons each? 
Whereas a larger focus can dramatically increase the odds of illuminating randomly injected particles, it also drastically decreases the number of scattered photons should a particle be illuminated ($N$).
These odds, also known as the `hit-rate', is effectively $\Mdata$ per time. 
In fact, $N\propto L_\text{focus}^{-2}$ while  $\Mdata/\text{time} \propto L_\text{focus}^{2}$.
In this hypothetical scenario, the total number of photons measured per time ($N \Mdata /\text{time}$) remains constant despite $L_\text{focus}$. 
Suppose that in either case, you had enough patterns to adequately sample different views of the scatterer, and were perfectly able to detect particle hits against background scatter/noise. 
This same ambivalence to the focus size appears again in the simple signal-to-noise ratio (SNR) described in \cite{Loh2009-eg}: 
\begin{equation}
    \text{SNR} = \left(\frac{N \Mdata }{M_{\text{rot}}}\right)^{1/2} \;, \label{eqn:SNR}
\end{equation}
where $M_{\text{rot}}$ is the number of rotation samples used to reconstruct the intensity volumes $W_A$ and $W_B$. This SNR is motivated by a simple distribution of photons across a limited number of Ewald tomograms, and has been used to indicate data sufficiency in the orientation space \cite{ayyer2016dragonfly}.

The discussion above may lead one to believe that there is no ideal focus size. 
However, if we again used a smaller orientation disconcurrence $\Delta \theta_\text{c}$ to quantify when things are `better', the preference is to reduce $L_\text{focus}$. 
Notice that nearly doubling the average number of photons per pattern ($N =355$ to $N=622$ given $\Mdata=5000$) in Fig.~\ref{fig:dtheta-photon} reduces both $\Delta \theta_\text{c}$ and $\Delta \theta_\text{i}$ more than if we doubled the number of patterns ($\Mdata=5000$ to $\Mdata=10000$ given $N=355$) in Fig.~\ref{fig:dtheta-pattern}. 
The total number of photons in all patterns is approximately equal in both cases. Yet doubling the average number of photons per pattern substantially improves the asymptotic orientation inconsistency (i.e.~$\Delta \theta_\text{i}^\ast$ falls).

\section{Discussion}
In summary, we propose an encryption-decryption approach to validate 3D intensity volumes reconstructed in XFEL-SPI. 
This validation is based on the volumes' ability to decrypt the orientations of sentinel patterns unused in these reconstructions.
While these volumes can be reconstructed from any algorithmic means, they must strictly adhere to the data independence scheme laid out in Fig. \ref{fig:cryptography}.
This scheme can be generalized to validate other latent information inferred within the full dataset (e.g.~unmeasured local photon fluence, structural class, etc). 

From realistic simulations of SPI experiments this approach can validate reconstructions in a principled information-theoretic manner. 
Our approach relates the challenging question of data sufficiency intuitively to key experimental variables such as the number of measured photon patterns, and nominal incident photon intensity.
Furthermore, the various forms of decrypting (orientation) uncertainties shown here can be interpreted as disconcurrence, disagreement, and inconsistencies in how confidently the latent variables are inferred. 
These interpretations give a more informative and comprehensive view of the validation exercise.

Whereas there were studies about the expected scattered photon signals from biomolecules in idealized XFEL-SPI scenarios \cite{Shen2004-cq, Giewekemeyer2019-ny}, there systematic studies of how well these signals can be integrated into a 3D diffraction volume despite missing information when is still sorely lacking.
Our results show that the complex considerations that contribute to data sufficiency in XFEL-SPI can be fitted as simple parameters (e.g.~$\alpha, \beta, \gamma$).
Relating these parameters to basic properties of the target scatterer (e.g.~mass, radius of gyration, etc), experimental conditions (e.g.~beam intensity, photon wavelength, background scattering, etc), and choice of reconstruction algorithms, will be useful for experiment design and planning. 

An extension of our encryption-decryption approach can be used to define and validate the spatial resolution of XFEL-SPI and cryo-electron microscopy reconstructions.
In principle, the resolving power of an imaging instrument should be the reduction in uncertainty of locating spatial features within the sample. 
Re-framing this uncertainty reduction in the encryption-decryption framework of Fig.~\ref{fig:cryptography} may give rise to more interpretable notions of spatial resolution.
This information theoretic formulation of this framework, similar to \myeqref{eqn:MutualInformationDecoding}, also naturally accounts for external priors for localizing spatial features.

Ultimately, our encryption-decryption approach demonstrably overcomes the difficulties of using FSC as a validation measure for XFEL-SPI, in spite of FSC's popularity\cite{hantkeHighthroughputImagingHeterogeneous2014,xuSingleshotThreedimensionalStructure2014,ekebergThreeDimensionalReconstructionGiant2015,
ayyerLowsignalLimitXray2019,giewekemeyerExperimental3DCoherent2019,hosseinizadehConformationalLandscapeVirus2017,
ikonnikovaReconstruction3DStructure2019,kimReconstruction3DImage2020,nakanoSingleparticleXFEL3D2018,
poudyalSingleparticleImagingXray2020,pryorSingleshot3DCoherent2018,roseSingleparticleImagingSymmetry2018,
shiEvaluationPerformanceClassification2019,vonardenneStructureDeterminationSingle2018}.
The data throughput from XFELS will rapidly increase because of higher pulse repetition rates \cite{Sobolev2020-mv}, and more efficient sample injection techniques. 
This trend inevitably creates a larger data load, which in turn increases our reliance on statistical techniques to assign confidence to {\it de novo} structural reconstructions.
Such confidence is especially important when imaging structural ensembles with considerable flexibilities, or other structural variations. 
Despite the specificity of our validation routine to orientations, the encryption-decryption framework proposed in Fig.~\ref{fig:cryptography} can be readily generalized to test the reproducibility of claims of novel reconstructed structures.
Such tests, we believe, are central to illuminating our path towards novel structural insights as we navigate through the photon-limited world of XFEL-SPI.

\section{Methods}
\label{sec:method}

\subsection{Sampling orientations.} 
A scatterer can take on an infinite number of possible 3D orientations. In practice these orientations $Q$ are discretely sampled to angular divisions smaller than the intrinsic angular precision of the patterns (see Section \ref{subsec:discuss_resolution}).
We adopt a quasi-uniform sampling scheme based on \cite{Loh2009-eg}, which adaptively refines the 600-cell polytope with refinement parameter $n$. 
In this scheme the number orientation samples scales like $n^3$, while their angular resolution increases like $1/n$.

\subsection{Orientation posterior distribution (OPD) of sentinel patterns.}
The orientation posterior distribution (OPD) of a particular sentinel pattern $K_\text{S}$ defines the probability of orienting it within a specific 3D diffraction volume $W$. 
This OPD, written here as $P(Q\given K_\text{S},W)$, can be inferred from the likelihood $P(K_\text{S}\given Q, W)$ using Bayes' theorem,
\begin{equation}
P(Q \given K_\text{S},W) \propto P(K_\text{S}\given Q,W ) \, P(Q) \quad ,\label{eqn:posterior-likelihood}
\end{equation}
where the prior distribution of orientations, $P(Q)$, is uniformly distributed unless the specimens have a known orientation bias.
Because the space of orientations is only quasi-uniformly sampled by unit quaternions in our discretization scheme, we replace $P(Q)$ with the numerically computed non-uniform weights $w(Q)$ \cite{ayyer2016dragonfly}.
Note that this OPD can be computed even if $\KS$ did not in fact originate from $W$: such a computation will naturally yield highly uncertain orientations of $\KS$. 

We presume the likelihood of detecting a sentinel pattern $\KS$ (comprising pixels indexed by $t$) from the Ewald tomogram at orientation $Q$ of volume $W$ (see Fig. \ref{fig:rotation_schematic}) assuming perfect detection absent background photon sources is 
\begin{equation}
P(\KS\given Q,W ) = \prod_{t \in \text{detector}} \frac{ \text{e}^{-W_{Q i}} \,W_{Q t}^{K_{\text{S}t}} }{K_{\text{S}t}!} \quad . \label{eqn:likelihood}
\end{equation}
This likelihood can be replaced if the true detection statistics departs from this Poissonian form.
 

Often the posterior and likelihood in \myeqref{eqn:posterior-likelihood} and \myeqref{eqn:likelihood} of a converged intensity volume is significant only for a relatively small set of orientations. 
For a given pattern $\KS$, we represent this set of {\it important orientations} by their corresponding important unit quaternions $\{\bbQ \given \KS\}$ (written in boldface). 
For computation efficiency, only the probability at $\{\bbQ \given \KS\}$ is recorded; those at other quaternions are safely set to zero. 

For sufficient orientation coverage, we require these important quaternions to capture at least 99\% of the total posterior distribution. 
To implement this, all patterns' posterior distributions are first sampled by a unit quaternion set $\{Q \given n\}$ with 600-cell quaternion sampling strategy \cite{Loh2009-eg} where $n$ is the sampling refinement level. 
Then we increase $n$ until the smallest set of important quaternions $\{\bbQ \given \KS,n\}_{\text{min}} \subset \{Q \given n\}$ that captures this total posterior distribution comprises at least 100 important quaternions:
\begin{equation}
\Big\langle\sum_{Q \in \{\bbQ \given \KS, n\}_{\text{min}}} P(Q \given \KS, W)\Big\rangle_{\KS} \geq 0.99 \; ,
\label{eqn:imptQuat}
\end{equation}
and the size of every $\KS$, $|\{\bbQ \given \KS, n\}_{\text{min}}| \geq 100$. 
To be concise, we omit the subscript $\cdot_\text{min}$ in subsequent formulae.

\subsection{Angular displacement distribution (ADD) between two reconstructed volumes.}

Returning to our cryptography analogy, our next step is to compare how two diffraction volumes decrypt the orientations of a set of sentinel patterns.
Three key considerations stand out here. 
First, the orientation of a noisy sentinel pattern is described by a probability distribution (i.e.~OPD) rather than a point estimate. 
Second, $W_A$ and $W_B$ would almost always differ by an overall mutual 3D rotation $Q_{BA}$ because each volume is typically randomly initialized to avoid reconstruction biases. 
Hence, the sentinel OPDs for $W_A$ and $W_B$ would also be displaced by $Q_{BA}$.
Third, we must average the OPDs for different sentinel patterns to obtain a robust estimate of the orientation disconcurrence between $W_A$ and $W_B$.
These considerations are captured in the {\it angular displacement distribution} (ADD) between $W_A$ and $W_B$.

The ADD for a single sentinel pattern $\KS$ can be defined as the outer product of its OPD given $W_A$ and $W_B$ on their respective important quaternions, 
\begin{align}
P(\bbQ_{BA} | K_\text{S}, W_A, W_B) & \propto P(\bbQ_{A}|K_\text{S}, W_A) P(\bbQ_{B} |K_\text{S}, W_B) \nonumber \\
& \propto P(\bbQ_{A}|K_\text{S}, W_A) P(\bbQ_{BA} \bbQ_{A} |K_\text{S}, W_B) \; , \label{eqn:ADD-single}
\end{align}
which is computed over the set of important unit quaternions. 
Here  $\bbQ_{BA} = \bbQ_B \bbQ_A^{-1} $ represents the possible relative orientations between the reconstructed volumes $W_A$ and $W_B$ over the two sets of important quaternions $\{\bbQ_A | \KS \}$ and $\{\bbQ_B | \KS \}$ as defined in \myeqref{eqn:imptQuat}. 
Since $\bbQ_{BA}$ depends on the sentinel pattern $K_\text{S}$, the ADD in \myeqref{eqn:ADD-single} may be different for different $\KS$. 
Averaging the ADD over all the set of sentinel patterns $\{ \KS \}$ we get 
\begin{align}
    P(\bbQ_{BA} |\{ K_\text{S}\}, W_A, W_B) \equiv \Big \langle P(\bbQ_{BA} | {K_\text{S}}, W_A, W_B) \Big \rangle_{\{K_\text{S}\}}\; .
    \label{eqn:ADD-all}
\end{align}

Given the noise in the diffraction patterns, we expect variations in the decrypted orientations of sentinel patterns.
To compute this variation, an average of an ADD must be established.
When the reconstructed volumes $W_A$ and $W_B$ are similar, the ADD of their many sentinel patterns tend to cluster around the average unit quaternion $\overline{Q}_{AB}$ in orientation space. 
This overall rotation  $\overline{Q}_{AB}$ is not a mere linear average of the unit quaternions that sample the ADD since this average may not have unit length and hence not correspond to a 3D spatial rotation.
To define $\overline{Q}_{AB}$, let us first consider the relative rotation between $\bbQ_{BA}$ and a presumptive average overall rotation $\widetilde{Q}$. This relative rotation can be written as a quaternion multiplication
\begin{align}
    \bbQ_{BA}^{-1} \, \widetilde{Q} &= \Big \{ \cos\left( \frac{\theta}{2} \right), \, \sin\left( \frac{\theta}{2} \right) \hat{\bvec n} \Big \} \, ,
\end{align} 
which is written here as a four-component vector; $\hat{\bvec n}$ and $\theta$ are respectively the axis and magnitude of this relative rotation.
The magnitude of this relative rotation, $\theta(\bbQ_{BA}, \widetilde{Q})$, vanishes as $\widetilde{Q}$ approaches $\bbQ_{BA}$. 

We define the average overall rotation $\overline{Q}_{BA}$ of an ADD between $W_A$ and $W_B$ as that which minimizes the average $\theta$ against all the rotation samples of the ADDs for the set of sentinel patterns. 
Specifically, the average overall rotation is defined as the unit quaternion that maximizes the angular variance $\Theta^2$:
\begin{align}
    \overline{Q}_{BA} &\equiv \argmin_{\widetilde{Q}} \Theta^2\bigl( \widetilde{Q} \given[\big]\{K_\text{S}\}, W_A, W_B \bigr) \, ,
    \label{eqn:OverallOrientation}
\end{align}
and the orientation disconcurrence is the minimum value of $\sqrt{\Theta^2}$:
\begin{align}
\Delta \theta_\text{c}(W_A, W_B) &\equiv \min_{\widetilde{Q}} \sqrt{\Theta^2\bigl( \widetilde{Q} \given[\big]\{K_\text{S}\}, W_A, W_B \bigr)}\nonumber \\
                                        &=\sqrt{\Theta^2(\overline{Q}_{BA} \given \{K_\text{S}\}, W_A, W_B)}\;,
\label{eqn:OrientationConcurrence}
\end{align}
where the angular variance is defined as
\begin{align}
    &\Theta^2\bigl(\widetilde{Q} \given[\big]\{K_\text{S}\}, W_A, W_B \bigr) =\nonumber \\
    &\left\langle \sum_{\{\bbQ_{BA} \given K_\text{S}\}} P(\bbQ_{BA} \given K_\text{S}, W_A, W_B) \, \theta^2(\bbQ_{BA}, \widetilde{Q}) \right\rangle_{\{K_\text{S}\}}\;. \label{eqn:Theta}
\end{align}
A special case here is when $W_A$ and $W_B$ are identical.
In this case, $\overline{Q}_{BA}=(1,0,0,0)$ which is the identity quaternion. 

\subsection{Resolving ambiguities from centro-symmetric diffraction volumes.}
\label{subsec:centrosym}
To obtain the most compact ADD (\myeqref{eqn:ADD-all}), we must eliminate trivial symmetries in the diffraction patterns that broaden the ADD. 
One such example is the centro-symmetry of 3D diffraction intensities from optically thin samples, whose scattering density distribution is effectively real-valued.
Consequently, at sufficiently low resolutions any two-dimensional diffraction pattern is similar to itself after a \SI{180}{\degree} in-plane rotation about the scattering experiment's optical axis ($\hat{z}$). 
Each such photon pattern $K$ should have similar posterior probabilities to occur at either rotation $Q$ or $Q Q_z$: 
\begin{align}
    P(Q\given K, W) \approx P(QQ_z\given K,W) \; ,
    \label{eqn:InplaneAmbiguity}
\end{align}
where the in-plane rotation about the $z$-axis is $Q_z = (0,0,0,1)$.
This two-fold ambiguity plus the fact that $Q_z$ is its own inverse, means that in ADD, the relative rotation $Q_{BA}$ or $Q_{BA}^{\prime} = Q_B \,Q_z \,(Q_A)^{-1}$ could occur in \myeqref{eqn:ADD-all}.
Hence, for each ADD sample we check the angular closeness of both $Q_{BA}$ and $Q_{BA}^{\prime}$ to the ADD's average unit quaternion $\overline{Q}_{BA}$, and keep the one that is closer.
This essentially replaces the $\theta$ expression in \myeqref{eqn:Theta}:
\begin{align}
    \theta^2(\bbQ_{BA}, \widetilde{Q}) \to \text{min}\{\theta^2(\bbQ_B\bbQ_A^{-1}, \widetilde{Q}), \theta^2(\bbQ_B Q_z \bbQ_A^{-1}, \widetilde{Q})\} \; . \label{eqn:ThetaWithInPlaneRotation}
\end{align}
\subsection{Discrete symmetries in the diffraction volume.}
\begin{figure}
    \centering
    \includegraphics[width=0.8\textwidth]{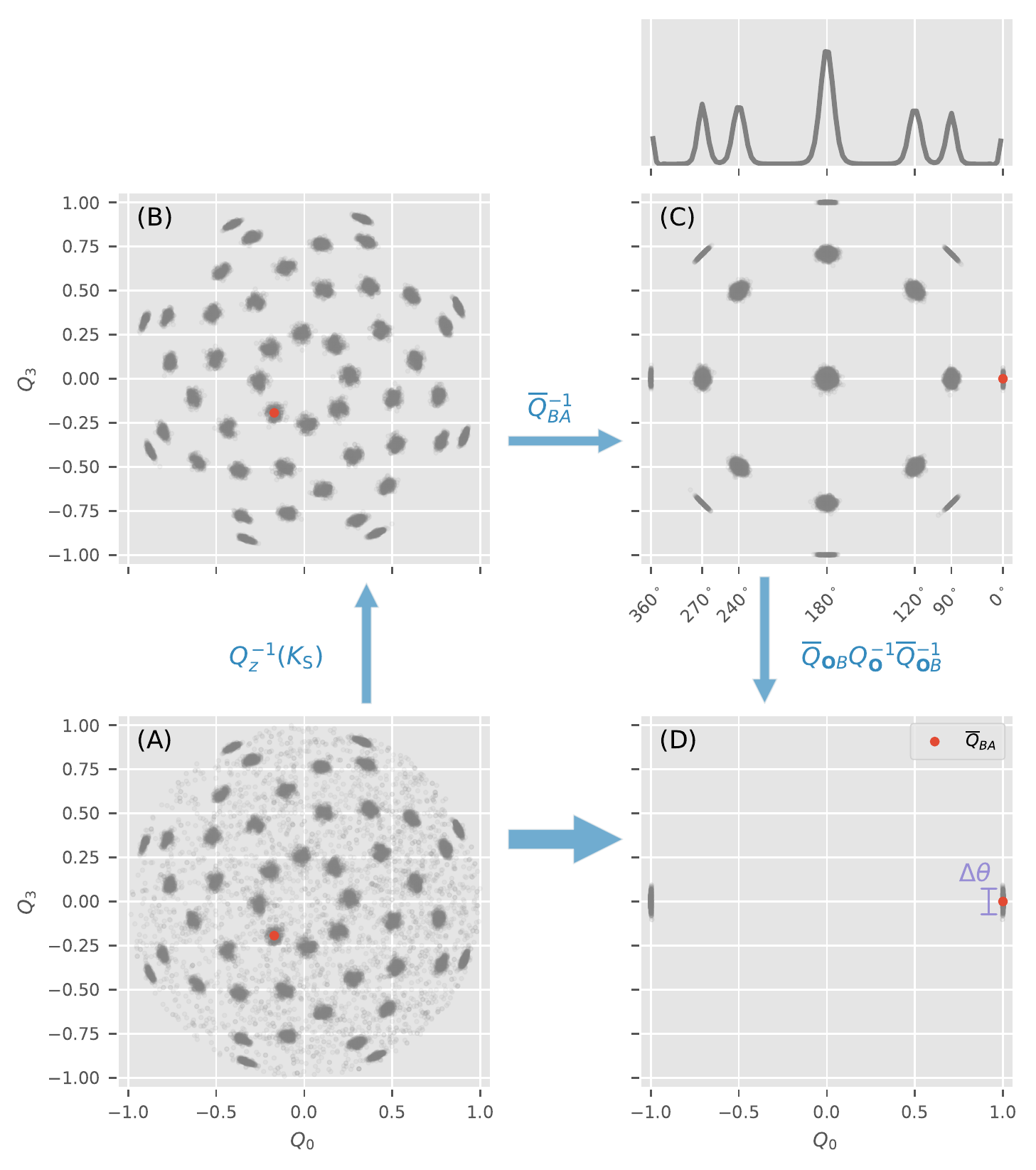}
    \caption{Collapsing the ADD of 500 sentinel patterns for a scatterer, whose diffraction volumes is centro-symmetric and has octahedral symmetry, into the fundamental domain: (A) to (D). Starting clockwise from (A), which shows a projection of the ADD onto two components of each quaternion ($Q = (Q_0, Q_1, Q_2, Q_3)$), we collapsed the points related by centro-symmetry (since 2D patterns have sufficiently low resolution) to obtain a sharper distribution in (B). The red disk throughout the panels represent the average quaternion $\overline{Q}_{AB}$ of the ADD. In (C), we rotate the ADD such that $\overline{Q}_{AB} = (1,0,0,0)$ for clarity. The histogram of the ADD vs $Q_0$ is shown above panel (C), can sometimes reveal the flavor of symmetry in $W$. Finally, using the particle's known symmetry group operations we can fold the ADD into the fundamental domain in (D).}
    \label{fig:oct_cluster}
\end{figure}
Discrete symmetries in the diffraction volume can create multiple clusters in the ADD (Fig.~\ref{fig:oct_cluster}). Examples of such symmetries include icosahedral viral capsids\cite{ekebergThreeDimensionalReconstructionGiant2015} and octahedral nanoparticles\cite{xuSingleshotThreedimensionalStructure2014}. 
The multiplicity of these clusters arise because each pattern could be oriented at different and/or multiple locations of the symmetry orbit within the diffraction volume. 
As Fig. \ref{fig:oct_cluster} shows, should this symmetry be known we can compute a single orientation disconcurrence by first folding these multiple symmetry-related peaks in ADD into its fundamental domain. 
We emphasize that this folding can be done even if this symmetry were not imposed during the reconstructions of $W_A$ and $W_B$.

Fig.~\ref{fig:oct_cluster} illustrates ADD folding for a particle with chiral octahedral symmetry ({\bf O}). The reconstructed diffraction intensities of this particle ($W_A$ and $W_B$) has 24 rotational symmetries (of order 24). Once $W_A$'s body axes are canonically aligned, then each of these symmetry rotations can be represented by a canonical set of unit quaternions $\{ Q_\text{\bf O} \given  \left[Q_\text{\bf O}\right] \in \text{\bf O}\}$ ($\left[Q_\text{\bf O}\right]$ is the equivalence class $Q_\text{\bf O} \sim -Q_\text{\bf O}$ owing to unit quaternions double covering SO(3)). 

To see how this symmetry manifests in an ADD, consider orienting a particular sentinel pattern $K_\text{S}$ within $W_A$ and $W_B$. Note that even though $W_A$ and $W_B$ have $\textbf{O}$ symmetry, they are not canonically aligned by default. First, we focus on  a tomogram of $W_B$ at $\bbQ_B$, $T(\bbQ_B, W_B)$. Here, the symbol for tomogram is changed from the $W_Q$ in the main text to avoid multiple level subscript. When we align $W_B$ canonically by actively rotating it to $\widetilde{Q}_{\mathbf{O}B}[W_B]$, the tomogram should be rotated together to maintain unchanged,
where $\widetilde{Q}_{\text{\bf O}B}$ actively rotates $W_B$ to $\widetilde{Q}_{\mathbf{O}B}[W_B]$  into the canonical axes for the symmetry operations in $\{Q_\text{\bf O}\}$.
In other words, we have
\begin{align}
    T(\bbQ_B, W_B) &= T\bigl(\widetilde{Q}_{\mathbf{O}B}\bbQ_B, \widetilde{Q}_{\mathbf{O}B}[W_B]\bigr)\\
    &=T\bigl(\widetilde{Q}_{\mathbf{O}B}\bbQ_B, (Q_\mathbf{O}\widetilde{Q}_{\mathbf{O}B})[W_B]\bigr)\\
    &=T\bigl(\widetilde{Q}_{\mathbf{O}B}^{-1}Q_\mathbf{O}^{-1}\widetilde{Q}_{\mathbf{O}B}\bbQ_B, W_B\bigr)\text{.}
\end{align}
The 24 elements in $\{Q_\text{\bf O}\}$ give 24 same tomograms at $\widetilde{Q}_{\mathbf{O}B}^{-1}Q_\mathbf{O}\widetilde{Q}_{\mathbf{O}B}\bbQ_B$ (all $Q_\mathbf{O}^{-1}\in \{Q_\mathbf{O}\}$ also), hence the same orientation posterior probability at these orientations.
Recalling the ADD comprises the joint product of OPDs for $K_\text{S}$ to be oriented at $\bbQ_A$ and $\bbQ_B$ within $W_A$ and $W_B$ respectively. 
We see this multiplicity of ADD in Fig.~\ref{fig:oct_cluster}b (main text), which contains 48 clusters owing to the the unit quaternion double covering $\text{SO}(3)$. 
The number of clusters does not increase even if we include the symmetry operations of $W_A$ by assuming $W_A$ and $W_B$ are similar, for the same reason that randomly oriented sentinel patterns in an asymmetric volume still produce a 2-clustered ADD (only one branch is plotted in Fig.~\ref{fig:4dcluster}). 

For each sentinel pattern $K_\text{S}$, we can fold each important unit quaternion $\bbQ_{BA}$ in its ADD into the fundamental domain by exhaustively searching the symmetry operation in $\bigr\{\widetilde{Q}_{\mathbf{O}B}^{-1}Q_\mathbf{O}\widetilde{Q}_{\mathbf{O}B}\bbQ_B\given[\big]Q_{\mathbf{O}}\in\{Q_\mathbf{O}\}\bigr\}$ and in-plane inversion $Q_z$ (either $\{1,0,0,0\}$ or $\{0,0,0,1\}$) that minimizes the angular variance
\begin{align}
    &\theta^2_\text{min}\left(\widetilde{Q}_{\text{\bf O}B}, \widetilde{Q} \given K_\text{S}, \bbQ_{BA}\right) = \nonumber \\
    &\min_{\{Q_\text{\bf O}\} \times \{Q_z\}} \theta^2\left(  \widetilde{Q}_{\text{\bf O}B}^{-1} Q_\text{\bf O}\, \widetilde{Q}_{\text{\bf O}B} \bbQ_B Q_z \bbQ_A^{-1}, \widetilde{Q} \given K_\text{S} \right)\; . \label{eqn:CheckSymmetryOrbit}
\end{align}
Here, $\widetilde{Q}$ is the presumptive average relative rotation between $W_A$ and $W_B$ similar to that in \myeqref{eqn:OverallOrientation}.
Like \myeqref{eqn:ThetaWithInPlaneRotation}, we also minimize over each pattern's in-plane inversion. 
Therefore, the optimal relative rotation ($\overline{Q}_{BA}$) and canonical realignment ($\overline{Q}_{\text{\bf O}B}$) are found by minimizing the total angular variance weighted over all important unit quaternions for all sentinel patterns in the ADD: 

\begin{align}
    &(\overline{Q}_{\text{\bf O}B}, \; \overline{Q}_{BA}) = \argmin_{(\widetilde{Q}_{\text{\bf O}B}, \; \widetilde{Q})} \Theta^2\left(\widetilde{Q}_{\text{\bf O}B}, \widetilde{Q} \given \{K_\text{S}\}, W_A, W_B \right)\;, \nonumber\\
    &\text{where} \nonumber \\
    &\Theta^2\left(\widetilde{Q}_{\text{\bf O}B}, \widetilde{Q} \given \{K_\text{S}\}, W_A, W_B \right) = \nonumber \\
    &\left\langle \sum_{\{\bbQ_{BA} \given K_\text{S}\}} P(\bbQ_{BA} | K_\text{S}, W_A, W_B)\, \theta^2_\text{min}\left(\widetilde{Q}_{\text{\bf O}B}, \widetilde{Q} \given K_\text{S}, \bbQ_{BA}\right) \right\rangle_{\{K_\text{S}\}} \;. \label{eqn:Canonicalization}
\end{align}

To recapitulate, the orientation disconcurrence between two symmetric volumes $W_A$ and $W_B$ is defined by \myeqref{eqn:Canonicalization} as 
\begin{equation}
   \Delta \theta_c^2 = \Theta^2\left(\overline{Q}_{\text{\bf O}B}, \overline{Q}_{BA} \given \{K_\text{S}\}, W_A, W_B \right)\; .\label{eqn:SymmetricOrientationConcurrence} 
\end{equation}
This computation involves separate optimizations: 
we iteratively refine $\widetilde{Q}_{BA} \to \overline{Q}_{BA}$ and $\widetilde{Q}_{\text{\bf O}B} \to \overline{Q}_{\text{\bf O}B}$ by minimizing \myeqref{eqn:Canonicalization}; 
for each presumptive $\widetilde{Q}_{BA}$ and $\widetilde{Q}_{\text{\bf O}B}$, find the symmetry operation in $\{Q_\text{\bf O}\}$ for each sentinel pattern that minimizes the quantity in \myeqref{eqn:CheckSymmetryOrbit} as well as the most compatible in-plane rotations for each sentinel pattern (Section~\ref{subsec:centrosym}).
The results of these completed optimizations are used to fold the ADD into the fundamental domain in Fig.~\ref{fig:oct_cluster}.

We note that one can discover the symmetry of $W_A$ using a special case of ADD with itself (i.e.~$W_A = W_B$). 
This `self-ADD' will be similar to Fig.~\ref{fig:oct_cluster}c (main text) since there is no relative rotation between $W_A$ and itself. 
Because the first component of every unit quaternions in a symmetry group is independent on the choice of canonical axis, we may deduce $W_A$'s symmetry group from number and positions of their clusters in their $Q_0$ histograms of its `self-ADD' (panel above Fig.~\ref{fig:oct_cluster}c (main text)).

\subsection{A one-dimensional (1D) model}
\label{appendix:1dmodel}
Here, we show the relation between the orientation disconcurrence and the disagreement (misalignment of the centers of ADDs) and the inconsistency (the size of each ADDs) with a one-dimensional (1D) rotation analogy as opposed to the full 3D rotation version in Fig.~\ref{fig:4dcluster}.

The unit quaternion $\bbQ$ that describes rotation about a 1D ring is a real number $\theta \in [0, 2\pi)$.
Suppose that the two OPDs (of reconstructed models $W_A$ and $W_B$) that comprise the ADDs for a set of sentinel patterns $\{\KS\}$ are mostly constrained within a small segment of this 1D ring. 
Let us further suppose that their ADD over $\{\KS\}$ can be approximated by local Gaussian distribution within this angular segment.
We denote the 1D ADD averaged over all sentinel patterns $\{\KS\}$ as $P(\bvec Q\given \{\KS\})\equiv P(\bvec Q\given \{\KS\}, W_A, W_B)$.
For a single sentinel pattern $K_\text{S}$ its ADD, $P(\bvec Q\given K_\text{S})$ (blue or red distribution in Fig.~\ref{fig:4dcluster}), we denote its mean as $\overline{Q}(K_\text{S})$, and variance as $\Delta\theta^2(K_\text{S})$.
Hence the mean and variance of this ADD for the entire set of sentinel patterns $\{\KS\}$ are equivalent to the overall orientation, $\overline{Q}(\{\KS\})$, and the square of orientation disconcurrence, $\Delta\theta_\text{c}^2(\{\KS\})$, defined in \myeqref{eqn:OrientationConcurrence} and \myeqref{eqn:Theta} respectively.
The square difference between the disconcurrence, $\Delta \theta_\text{c}(\{K_\text{S}\})$, and 
the inconsistency, $\sqrt{\braket{\Delta \theta^2 (K)}_{K\in\{K_\text{S}\}}}$,
is equivalent to the RMS distance between $\overline{Q}(\KS), \KS\in\{\KS\}$ and $\overline{Q}(\{\KS\})$, can be thought of as the disagreement, $\Delta\theta_\text{a}(W_A,W_B)$, between reconstructions $W_A$ and $W_B$. This relation can be shown by
\begin{equation}
\begin{aligned}
   &|\{K_\text{S}\}|\Delta \theta_\text{c}^2(\{K_\text{S}\}) - \sum_{\KS}\Delta \theta^2(K_\text{S}) \\
   =&\sum_{\KS}\sum_{\bvec Q} P(\bvec Q\given \KS)\big(\bvec Q - \overline{Q}(\{\KS\})\big)^2\\
   &-\sum_{\KS}\sum_{\bvec Q} P(\bvec Q\given \KS)\big(\bvec Q - \overline{Q}(\KS)\big)^2\\
   =&\sum_{\KS}\sum_{\bvec Q} P(\bvec Q\given \KS)\big(\bvec Q^2 - 2\bvec Q \overline{Q}(\{\KS\})+\\
   &\overline{Q}^2(\{\KS\}) - \bvec Q^2 +2\bvec Q\overline{Q}(\KS)-\overline{Q}^2(\KS)\big)\\
   =&\sum_{\KS}\sum_{\bvec Q} P(\bvec Q\given \KS)\big(- 2\overline{Q}(\KS) \overline{Q}(\{\KS\})+\\
   &\overline{Q}^2(\{\KS\}) +2\overline{Q}(\KS)\overline{Q}(\KS)-\overline{Q}^2(\KS)\big)\\
   =&\sum_{\KS}\sum_{\bvec Q} P(\bvec Q\given K)\big(\overline{Q}(\KS) - \overline{Q}(\{\KS\})\big)^2\\
   =&\sum_{\KS}\big(\overline{Q}(\KS) - \overline{Q}(\{\KS\})\big)^2\\
   \equiv& \Delta\theta_\text{a}(W_A,W_B)\text{.}
\end{aligned}
\label{eq:1d_consis_agree}
\end{equation}
Above we use $\sqrt{\braket{\Delta \theta^2 (K)}_{K\in\{K_\text{S}\}}}$ as the inconsistency in \myeqref{eq:1d_consis_agree} instead of the definition in \myeqref{eqn:OrientationConsistency}, because these two definitions are approximately
the same if Gaussian distributions are assumed for OPDs, $P(\bbQ_i\given K_\text{S}, W_i)$, $i=A, B$. 
As $P(\bvec Q \given K_\text{S})$ is a convolution of these two Gaussian OPDs, its variance is $\Delta\theta^2(K_\text{S})=\delta_A^2 + \delta_B^2$, where
$\delta_A^2$ and $\delta_B^2$ are the variances of $\text{OPD}_A$ and $\text{OPD}_B$.
Meanwhile, the variances of auto-convolution of two OPDs are $\Theta^2(\overline{Q}_{ii}=0 \given K_\text{S}, W_i)=2\delta_i^2$, $i=A, B$, which gives us
\begin{equation}
    \Delta\theta^2(K_\text{S}\given W_A, W_B) \approx \frac{1}{2} \Theta^2(0\given K_\text{S}, W_A) +
    \frac{1}{2} \Theta^2(0 \given K_\text{S}, W_B)=\Delta\theta_\text{i}^2(W_A, W_B)\text{.}
    \label{eqn:1D_delta_split}
\end{equation}
The average of right hand side (RHS) of \myeqref{eqn:1D_delta_split} over $\{K_\text{S}\}$ is consistent with RHS of \myeqref{eqn:OrientationConsistency}.

The width of OPD, $\delta^2$, quantifies how well we can identify the orientation for a given pattern. For a pixel at $\bvec q$ in this pattern, we cannot decide whether this pixel belongs to a diffraction speckle near its most likely orientation if the speckle's radii $\theta_\text{sp}(\bvec q)$ is larger than $\delta$. Strictly, if we want a $74\%$ confidence interval, then we should have $\theta_\text{sp}(\bvec q) \leq 2 \delta$. 
It should be noted that the confidence interval for $2\sigma$ is $74\%$ instead of $95\%$ since OPD is a 3D Gaussian distribution even though we simplified the derivation above with a 1D Gaussian distribtuion.  
The $\delta$ is computational expensive, but it can be easily inferred from $\Delta\theta_\text{i}$ by $\delta\approx\Delta\theta_\text{i} / \sqrt{2}$ if the Gaussian assumption discussed above is utilized. Moreover, being more cautious about the conclusion,
we replace the $\Delta\theta_\text{c}$ instead of $\Delta\theta_\text{i}$ in \myeqref{eqn:resolutionCondition}.

\bibliographystyle{unsrt}
\bibliography{ref.bib}
\end{document}


\maketitle
\section{Representing spatial rotation with quaternions} \label{appendix:quaternions}
In this section, a brief introduction is given about the unit quaternion representation of rotation, which commonly occurs in computational geometry.
Quaternions are points in a 4D real space $(Q_0, Q_1, Q_2, Q_3)$. 
And the unit quaternion ($\sum_{i=0}^3 Q_i^2 = 1$) representation of a rotation has the following relation with the angle-axis pair representation, $(\theta\in [0, 2\pi), \hat{\bvec n})$,
\begin{equation}
Q = (\cos \frac{\theta}{2}, \sin \frac{\theta}{2}\cdot \hat{\bvec n})\text{,}
\label{eq:relation_quat_angleaxis}
\end{equation}
where $\hat{\bvec n}$
is the axis of the rotation and $\theta$ is the rotation angle.

The combination of two rotations is mapped to a special multiplication, which makes quaternions into an algebra. To define that multiplication, it is easier to rewrite \myeqref{eq:relation_quat_angleaxis} into $Q = Q_0 \bvec 1 + Q_1 \bvec i + Q_2 \bvec j + Q_3 \bvec k$, where $\bvec 1$, $\bvec i$, $\bvec j$, $\bvec k$ are four basis unit quaternions $(1, 0,0,0)$, $(0, 1, 0, 0)$, $(0, 0, 1, 0)$, $(0, 0, 0, 1)$.
The multiplication between any two quaternions can be extended from the multiplication table (Table ~\ref{tab:multitable}) of these four bases.
\begin{table}
    \caption{Multiplication table of basis unit quaternions}
    \label{tab:multitable}
    \centering
    \begin{tabular}{c|cccc}
        \hline
        & $\bvec 1$ & $\bvec i$ & $\bvec j$ & $\bvec k$\\ \hline
        $\bvec 1$ & $\bvec 1$ & $\bvec i$ & $\bvec j$ & $\bvec k$\\
        $\bvec i$ & $\bvec i$ & $-\bvec 1$ & $\bvec k$ & $-\bvec j$\\
        $\bvec j$ & $\bvec j$ & $-\bvec k$ & $-\bvec 1$ & $\bvec i$\\
        $\bvec k$ & $\bvec k$ & $\bvec j$ & $-\bvec i$ & $-\bvec 1$\\
        \hline
    \end{tabular}
\end{table}

Simply applying this bases multiplication shows that the conjugate, $Q^* \equiv Q_0 \bvec 1 - Q_1 \bvec i - Q_2 \bvec j - Q_3 \bvec k$, of a unit quaternion $Q$ is its inverse, $Q Q^* = \bvec 1$. This conclusion also can be verified by considering the quaternion representation of a rotation $(\theta, \hat{\bvec n})$ and its inverse $(\theta, -\hat{\bvec n})$.
Another helpful corollary derived from the multiplication is about
calculating the natural geodesic distance between two rotations, $\Omega_A, \Omega_B\in\text{SO}(3)$.
This distance is defined as the angle of rotation, $\theta$, of the joint rotation operation
$\Omega_A\cdot\Omega_B^{-1}$, or $\theta(Q_A, Q_B) = 2\cdot\arccos
\sum_{i=0}^{3} Q_{Ai}Q_{Bi}$ in the quaternion representation.

It should be noted that the positions of the OPD clusters in Fig.~\mref{fig:oct_cluster} are centro-symmetric. 
The reason is that rotating an object by $\theta$ along axis
$\hat{\bvec n}$, could also be expressed as rotating it by
$2\pi-\theta$ along axis $-\hat{\bvec n}$. 
However, the quaternions representations of these two equivalent rotations, $(\theta, \hat{\bvec n})\rightarrow Q$ and
$(2\pi-\theta, -\hat{\bvec n}) \rightarrow -Q$, are different by
\myeqref{eq:relation_quat_angleaxis}. 
Hence, we call the unit quaternion representation a {\it double cover} of $\text{SO}(3)$ group (also known as the 3D rotation group).